\documentclass[conference]{IEEEtran}
\usepackage{graphicx}
\usepackage{subcaption}
\usepackage{amsmath,amssymb,epsfig,xcolor} 
\usepackage{tabularx,multirow,array}
\usepackage{times}
\usepackage{booktabs}
\usepackage{multirow}
\usepackage{soul}
\usepackage{enumerate}
\usepackage[countmax]{subfloat}
\usepackage{algpseudocode}
\usepackage{algorithm}
\usepackage{threeparttable}
\usepackage{cite}
\usepackage{url}
\begin{document}
\title{On Detecting and Preventing Jamming Attacks with Machine Learning in Optical Networks }
\author{
    \IEEEauthorblockN{Mounir Bensalem, Sandeep Kumar~Singh, and Admela~Jukan}
    \IEEEauthorblockA{Technische Universit\"at Braunschweig, Germany}
     \IEEEauthorblockA{\{m.bensalem, sandeep.singh, a.jukan\}@tu-bs.de}
}
\maketitle
\begin{abstract}
Optical networks are prone to power jamming attacks intending service disruption. This paper presents a Machine Learning (ML) framework for detection and prevention of jamming attacks in optical networks. We evaluate various ML classifiers for detecting out-of-band jamming attacks with varying intensities. Numerical results show that artificial neural network is the fastest ($10^6$ detection per second) for inference and most accurate ($\approx 100 \%$) in detecting power jamming attacks as well as identifying the optical channels attacked. We also discuss and study a novel prevention mechanism when the system is under active jamming attacks. For this scenario, we propose a novel resource reallocation scheme that utilizes the statistical information of attack detection accuracy to lower the probability of successful jamming of lightpaths while minimizing lightpaths' reallocations. Simulation results show that the likelihood of jamming a lightpath reduces with increasing detection accuracy, and localization reduces the number of reallocations required.
\end{abstract}

\section{Introduction}
\par Optical Networks have long been known as vulnerable to different types of attacks, including power jamming attacks \cite{medard1998attack}. The jamming attacks are usually launched by inserting a relatively high power signal over either a frequency within the transmission window (\emph{in-band jamming}) or out of the transmission band (\emph{out-of-band jamming}) of legitimate data channels. Both types of jamming are common and can disrupt communication or even steal information with the help of crosstalk mechanism. The out-of-band jamming can be harder to detect under some circumstances, where the power of the jamming channel does not differ much than the legitimate channels. Furthermore, an out-of-band jamming signal at an erbium-doped amplifier (EDFA) can cause gain competition and starve legitimate signals from amplification. When a relatively high power signal  passes through multiple EDFAs along its route in wavelength-division multiplexed (WDM) networks, it accumulates higher amplified spontaneous emission noise and fiber nonlinear interference noise. These phenomena lead to signal degradation in terms of optical signal-to-noise ratio (OSNR) and bit error rate (BER) \cite{huang2017dynamic}. 

\par Previous work proposed attack prevention mechanisms based on data encryption, quantum key distribution, and scrambling techniques \cite{prucnal2009physical, hu2015chaos,li2016fast,singh2017combined}. While these methods are hard to break, attackers can still try to launch signal insertion and splitting attacks without being noticed. Therefore, an attack detection -- in addition to prevention itself -- is fundamentally important as well. Currently, network operators rely on power detection, spectrum analysis,  reflectometry methods and manual analysis by technicians for the jamming detection. But given the heterogeneity in network devices and technologies, these methods are prone to errors under noisy condition and may lead to low detection rates or prolonged detection delays \cite{chen2019self}. With increasing data rate of today's cloud-centric optical networks, attack detection, localization (i.e., identification of an attacked channel) in real time are highly desired but also challenging. To this end, machine learning (ML) has recently gained significant attention as a statistical method and a viable solution to attack detection to address the issues of attack detection and prevention in real time and accurately \cite{li2018light}. \cite{natalino2018field} applied support vector machine as a ML technique to detect jamming with 100\% accuracy under a strong attack. However, a detailed study of the effectiveness of various ML solutions to detecting jamming attacks under a wider range of jamming intensity is still missing. Equally important problems to be addressed are attack localization and prevention. 

\par This paper, therefore, studies the effectiveness of various ML solutions to detection, localization and prevention of power jamming attacks in optical networks. For detection and localization of out-of-band power jamming attacks in WDM networks, we first comparatively study the performance of various ML techniques, such as artificial neural networks (ANN), support vector machine (SVM), logistic regression (LR), K-nearest neighbors (KNN), decision tree (DT) and Naive Bayesian (NB). We also discuss jamming attack prevention and study the scenario of an active jamming attack. We show that after detecting an attack, reallocation of routes and wavelengths to all lighpaths leads to low probability of continuous and future jamming or observing of legitimate data signals carried by authorized lightpaths (LP). The challenge with the reallocation scheme is to use a few reconfigurations (through resource reallocation process) as possible, so that the services do not get unnecessarily disrupted. Thus, utilizing the attack localization feature, we reallocate resources to those lighpaths that are within the influence (attack radius) of jamming channels. Our analysis shows that a single reconfiguration in a LP's lifetime or around 8\% reconfigurations are sufficient to secure LPs, while ANN and SVM are the most suitable for attack detection and localization. 

\par The rest of this paper is organized as follows. Section \ref{sec:Model} presents a ML framework for attack detection and prevention.  Section \ref{sec:results} presents data extraction method for a WDM network scenario, attack prevention mechanism,  and performance evaluation. Finally, we conclude  the paper in Section \ref{sec:conclusion}.
\begin{figure*}[ht!]
 \centering
 \includegraphics[width=0.95\textwidth]{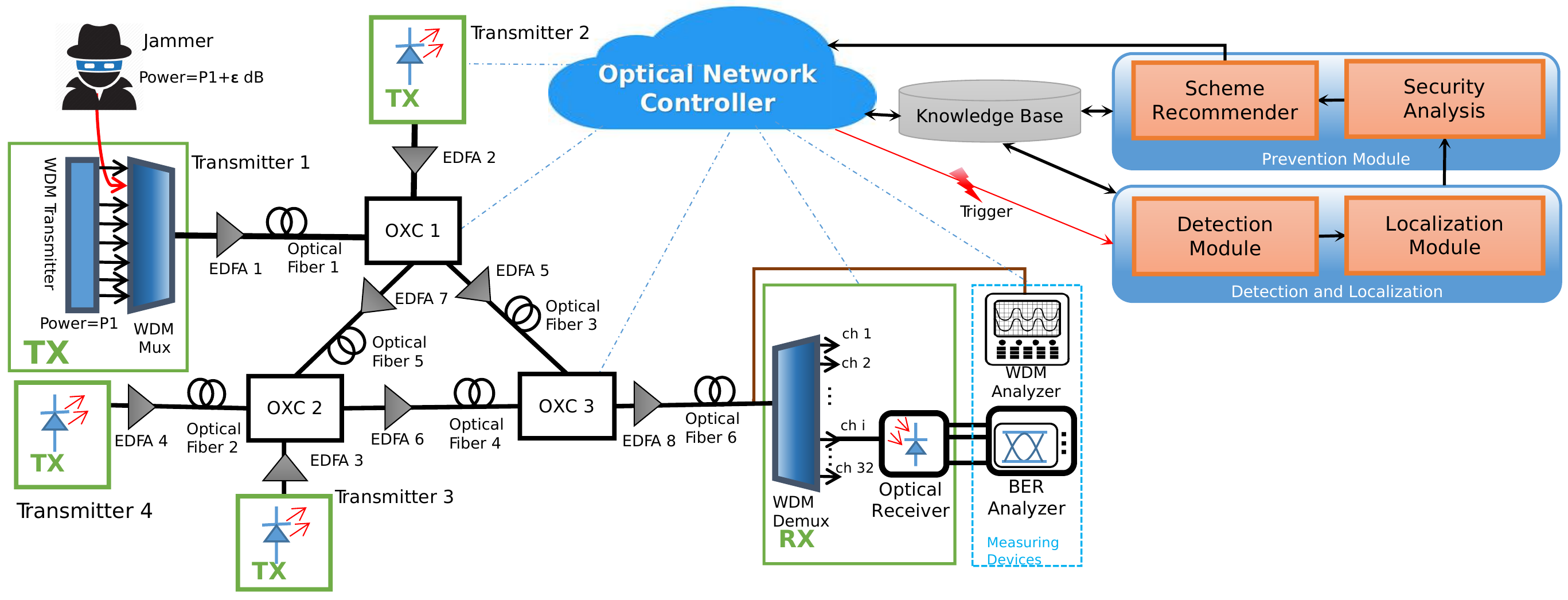}
  \vspace{-0.3 cm}
 \caption{A jamming attack detection and prevention framework. 
 }
\label{fig:networkarch}
\vspace{-0.4 cm}
\end{figure*}

\section{ML-based Attack Detection and Prevention} \label{sec:Model}
\setlength{\textfloatsep}{0.3cm} 
\setlength{\floatsep}{0.3cm}
The proposed ML framework for detection and prevention of jamming attacks in optical networks is shown in  Fig. \ref{fig:networkarch}. The framework consists of an optical network under a jamming attack, a network controller to monitor and reconfigure physical devices, and attack detection and prevention modules. The network controller collects optical performance monitoring data from transceivers and switches, and stores in a \textit{Knowledge Base}. The controller could trigger an attack detection procedure regularly or reactively upon detecting new/suspicions data. A \textit{Detection Module} then applies different ML techniques for detection of jamming attacks, and in case of an attack it triggers a \textit{Localization Module} to identify the jammed channel using a localization algorithm.  It is important to mention that the \textit{Localization Module} is generally designed to find the position of the attacker, however a network-wide attack monitoring is generally complex, and exact locations of  attacks might be difficult to identify in real networks. Thus, the studied localization is limited to the jammed channel. Using the attack detection and localization information, the state of network security can be analyzed, and  the controller can apply the best security appliance proposed by a \textit{Scheme Recommender} for attack prevention.  

\subsection{Jamming Attack Model }
The out-of-band jamming attacks in WDM networks generally require physical access to legitimate network access points, such as transmitter, coupler, optical cross-connect (OXC), etc. For simplicity, we define the power jamming attack as the ability of an unauthorized user to sneak into the network and access a light source (laser) to launch a signal with higher power than other authorized wavelength channels, targeting nonlinearity in the fiber and degradation of OSNR and BER of other neighboring channels. We distinguish between two types of user: authorized users, characterized by transmitter sources which generate optical signals with a defined signal power, and unauthorized users (jammers), characterized by transmitter sources with higher power signals. 

\par Fig. \ref{fig:networkarch} also illustrates our jamming attack model that we simulated to evaluate a jamming attack, as well as its detection and prevention in an optical WDM network with three OXCs, four transmitters and a WDM receiver.  Transmitter 1 illustrates authorized users and an unauthorized user in a WDM system, where the authorized sources in a WDM transmitter send  signals with power $P_1$, and an unauthorized signal is inserted by a Jammer in the $2^{nd}$ channel with $\varepsilon$ dB higher power than the legitimate WDM channels. As shown in Fig. \ref{fig:networkarch}, signals are multiplexed using a WDM Mux, then amplified with an  EDFA during the transmission before switched by an OXC to another fiber on its route. At the end, optical receivers and BER Analyzers can be used for monitoring of received signals. A WDM Analyzer is connected before the WDM Demux  to capture different features of all channels, e.g., signal power, noise power, SNR or OSNR. These features are analyzed to detect whether there is a jamming attack in a network using a detection algorithm, and an attacked channel is identified using a localization algorithm.

\subsection{Detection and Localization Formulation with ML}
The jamming detection/localization in optical networks can be formulated as a classification problem, which can be solved by typical supervised ML classifiers, such as ANN, DT, KNN, SVM, LR and NB. Therefore, we compare and evaluate these classifiers under a network scenario to find the most accurate technique for the attack detection framework. In order to use ML for jamming detection, we need to first collect data for training of ML classifiers. The data represent instances of a network where a jamming attack could happen. For each instance, an input represents measured features and output corresponds to a binary variable indicating the presence/absence of an attack.  We collect the information about two features: received channel power and OSNR. We monitor theses features by a WDM Analyzer.

Let's assume that there are  $T$ WDM transmitters in a network. An $i^{th}, 1 \leqslant i \leqslant T$ transmitter could establish $W$ number of LPs to a single WDM receiver over wavelengths $\lambda_{1}^i \leqslant \lambda_{w}^i \leqslant \lambda_W^i$ with equivalent channel power $P_w^i = P^i, 1 \leqslant w \leqslant W$. Let's assume that a Jammer $J$ can insert an unauthorized signal over any of the $T\times W$ unused wavelengths with a power $P^J = P^i+\epsilon$, where $1 \text{ dB } \leqslant \epsilon \leqslant 3$ dB. 
A received signal before demultiplexing corresponding to a wavelength channel $\lambda_{w}^i$ is denoted by $RS_{\lambda_{w}^i}=(f_{\lambda_{w}^i}^1,...,f_{\lambda_{w}^i}^m)$, where $f_{\lambda_{w}^i}^m$ is  the $m^{th}$ feature of the received signal over the channel $\lambda_{w}^i$.  In our study, we consider received signal power and OSNR of each channel as features for the detection and localization modules. Let's consider that we have multiple instances ($N_I$) of the network in Fig \ref{fig:networkarch}. An instance of the network generates one data point $n$ where $X_n = (RS_1,...,RS_{T\times W})$ is the input data. For the \textit{Detection Module}  the output $Y_n = 1$ if there is a jamming and $0$ otherwise, and for the \textit{Localization Module} the output $Y_n = \lambda_{w}^i$ if the channel $\lambda_{w}^i$ is jammed and $Y_n = 0$ if there is no attack. It is important to mention that  various ML techniques utilized for detecting jamming attacks use the same dataset consisting of input samples, each having a feature vector with length $T\times W$, i.e., $X_n \in \mathbb{R}^{T \times W}, 1\leqslant n \leqslant N_I$. 

\vspace{-0.2cm}
 \subsection{Discussion on Attack Prevention}
The behavior of jamming attacks depends on the goal of an attacker and his/her ability to access the network, which is hard to predict. Proposing only one prevention scheme can work in a specific scenario for which the scheme is designed and fail for other scenarios. A jammer can be smart and aware of the security system by using intelligent attack strategy: jamming with interruption, periodically or random attacks, or varying attack location. Another design problem is related to the complexity of an attack prevention scheme in terms of time and cost. Being aware of these dynamic problems, the \textit{Scheme Recommender} can select an appropriate prevention scheme based on the attacker's location and the history of attacks stored in the \textit{Knowledge Base}. In a general case, the \textit{Scheme Recommender} recommend the best security appliance using information provided by the \textit{Security Analysis}. Studying the behavior of a jammer and analyzing different security scenarios is out of the scope of this paper. Here, we mainly discuss a preliminary security scenario in which a network is under an active jamming attack without knowing the position of an attacker. We aim to minimize the number of channels affected by an attack. To this end, we propose to periodically change the optical resources allocated to maintain optical channels between a source-destination pair, called as lightpaths (LPs), using a resource reallocation strategy. The proposed attack prevention scheme is a proactive approach, in which the network controller can inquire about the network status regularly with mean interarrival time $\tau_r$, and as a response the attack detection and localization tool sends the likelihood of a network being attacked. The attack prevention scheme utilizes only the attack detection and localization probabilities, and based on these information it tries to reallocate optical resources to all or only a few LPs, such that an attacker or a group of attackers is unable to jam or have continuous access to authorized LPs. 

\section{Numerical Analysis} \label{sec:results}
We use an Optisystem software to simulate a jamming attack, and generate experimental datasets, which are used for training and testing of ML algorithms for detecting and localizing a jamming attack in a WDM network  scenario shown in Fig. \ref{fig:networkarch}. There are in total 32 optical channels from four WDM transmitters to a single WDM receiver over their shortest fiber-level paths. In other words, each WDM transmitter establishes 8 LPs.  For the simulation, we replace OXCs with power combiners, since the input signals (e.g., from Transmitter 1 and 2) at an OXC are destined for the same destination and traverses over the same fiber-level path thereafter. Thus, Fiber link 5 is removed for the simulation. The starting channel frequency at Transmitter 1 is 193.1 THz and the channel spacing is equal to 100 GHz. Each  WDM transmitter is connected to a WDM Mux with 1 dB insertion loss. The power of the multiplexed LPs coming from the first and fourth WDM transmitters is amplified with  EDFAs of 5 m length, and two EDFAs of 2 m are used to amplify the power of the LPs coming from the second and third WDM transmitters. We set the noise center frequency of each EDFA to the frequency corresponding to central channel of the multiplexed LPs. Thereafter, the  amplified LPs coming from EDFA 1 and EDFA 4 are connected to an optical fiber of 50 km length with 0.28 dB/km attenuation loss. The light from the Fiber 1 is combined with the amplified LPs from EDFA 2 using a power combiner. Similarly, we combine the LPs coming from fiber 2 and EDFA 3. After each combiner, the LPs are amplified using an EDFA of 5 m length then they travel a 70 km optical fiber. Afterward, we combine all LPs with a power combiner to get 32 multiplexed LPs. Finally, we amplify the LPs which traverse over an optical fiber of 50 km length before being demultiplexed by a WDM Demux and connected to separate optical receivers. A WDM Analyzer is used to extract the data related to each channel before the WDM Demux. We designate an optical transmitter source as a Jammer, which inserts an unauthorized signal with higher power. Moreover, the Jammer can select one of the 32 light sources in order to increase localization complexity.  \\
The power of each transmitter is set to a value between -22 and 0 dBm so that the received power of channels are close to each others, i.e., more power is given to such LPs that travel longer paths (relative to the 1st and 4th transmitters). The power of transmitters is varied in the defined range (-22 and 0 dBm) to creating enough and diverse data samples for the evaluation of ML algorithms. We use an Optisystem simulator to simulate various instances of authorized and unauthorized data samples, where each instance of generating a data sample takes around 20 to 25 minutes.

\vspace{-0.1cm}
\subsection{Jamming Attack Detection}
For evaluating the jamming detection capability without localization, the transmission power is varied for generating 3 types of datasets that store different percentage of authorized and unauthorized samples:  (50\%--50\%), (70\%--30\%) and (90\%--10\%). We analyze the detection on different datasets with different proportions of unauthorized samples to understand the effect of the jamming frequency on the detection accuracy. The unauthorized samples are generated by inserting higher power in one of the following channels: channel number 7, 13 and 27. The number of data points in the (50\%--50\%) dataset is 1140. As explained in Section \ref{sec:Model}, each input  sample holds the received power and OSNR information of all 32 channels. Each dataset is divided into training and testing sets with proportion of 75\% and 25\% respectively. 

We first evaluate the overall accuracy of different detection algorithms (ANN, DT, KNN, SVM, LR and NB) for the \textit{Detection Module} and analyze its two main components: True Positive (TP) rate as the probability of correct detection of jamming attacks, and True Negative (TN) rate as the right recognition of the normal behavior. We present the accuracy of each ML algorithm by averaging testing results with 20 different seeds. All algorithms are implemented in Python and use Scikit-learn library. The ANN is fully connected with one hidden layer, increasing hidden layers does not improve accuracy, and  a well-known L-BFGS is used as an optimizer. We consider only one neighbor for KNN, increasing k-neighbors decreased the accuracy, and the inverse of regularization strength is set to 0.1 in LR. Moreover, we considered the Gaussian NB, and  use the default parameters provided by scikit-learn implementation of SVM, LR and DT. 

\begin{figure}
    \centering
    \begin{subfigure}[b]{0.45\textwidth}
    \centering
        \includegraphics[width=0.9\textwidth]{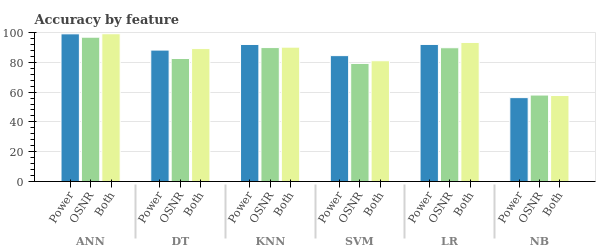}
        \label{fig:Accuracy}
    \end{subfigure}
    ~ 
    \begin{subfigure}[b]{0.45\textwidth}
    \centering
        \includegraphics[width=0.9\textwidth]{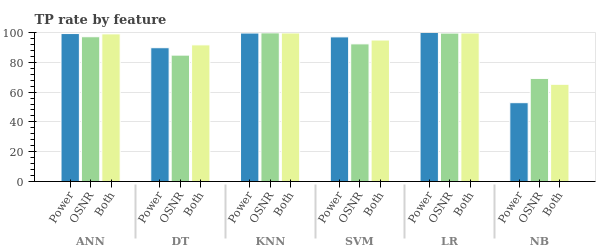}
        \label{fig:TP rate}
    \end{subfigure}
    ~ 
    \begin{subfigure}[b]{0.45\textwidth}
    \centering 
        \includegraphics[width=0.9\textwidth]{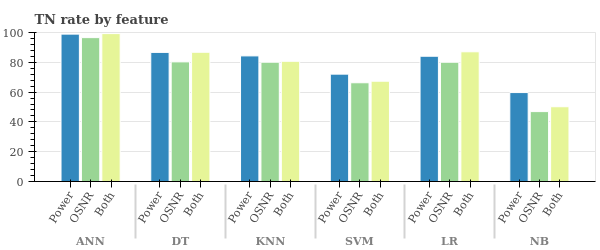}
        \label{fig:TN rate}
    \end{subfigure}
    \caption{Detection accuracy per alg. for 50-50\% sampling.}    
     \label{fig:accuracy5050}
     \vspace{-0.2cm}
\end{figure}
Fig. \ref{fig:accuracy5050} depicts the performance of different algorithms in terms of accuracy, TP rate and TN rate with equal proportion of authorized and unauthorized data samples, i.e., 50-50\% sampling. Except SVM, the accuracy under other algorithms is nearly same when both features are used rather than the power alone. The reason is that the jamming attack is realized by inserting a relatively higher power. Therefore, the received signal power is enough to model the jamming attack in optical networks. When we use the signal power alone as a feature, we see in Fig. \ref{fig:accuracy5050} that ANN outperforms other algorithms with an accuracy  of 99.1\%. LR, KNN, SVM and DT show an accuracy around 90\%. The NB technique is the worst candidate for detecting jamming attacks  with an accuracy of 56\% which means that the result is statistically analogous to a random selection in the case of a binary classification, i.e. 50\% to select $1$ and  50\% to select $0$.  

\begin{figure}[t]
 \centering
    \includegraphics[width=0.45\textwidth]{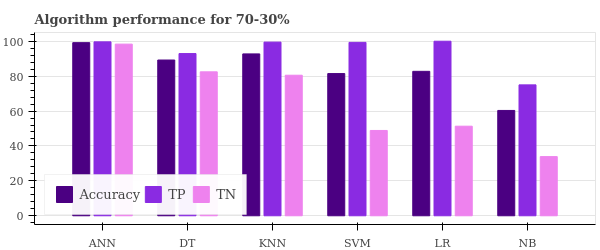}
    \includegraphics[width=0.45\textwidth]{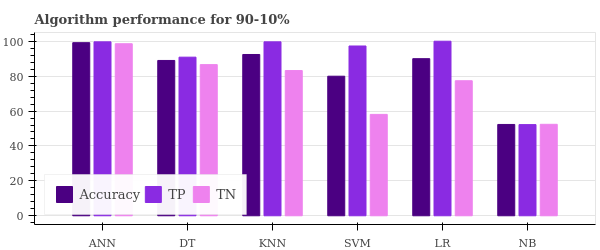}
  \vspace{-0.1 cm}
 \caption{Performance of algorithms with different sampling.}
\label{fig:algoPerformance}
\end{figure}
In Fig. \ref{fig:algoPerformance} we compare the performance of ML algorithms  considering two other datasets of 70-30\% and 90-10\% sampling, where the received power is used as the only feature.  We observe that ANN outperforms all other algorithms in different scenarios with an accuracy equal to 100\%. NB is still the worst technique for the detection problem in terms of accuracy. The accuracy of all other algorithms (DT, KNN, SVM and  LR) is between 80\% and 92\%, and their TN rates  decrease for 70-30\% and 90-10 \%. For SVM, the TN rate is equal to 48.9\% for 70-30\% sampling and 58.1\% for 90-10\% sampling. In general,  ANN is the only algorithm in the evaluated scenario that has a stable performance with different sampling strategies. For other algorithms, the TN rate is lower than 86\% in 90-10\% sampling. 
Furthermore, from Figs. \ref{fig:accuracy5050} and \ref{fig:algoPerformance} we observe that the accuracy of ML algorithms could increase when  the amount of authorized samples is increased in the dataset. The reason is that the TP rate increases  with increasing proportions of authorized samples (from 50\% to 90\%). On the other hand, the TN rate decreases sharply for most of ML algorithms when the percentage of unauthorized signals is decreased in the dataset 
\begin{figure}[t]
 \centering
    \includegraphics[width=0.45\textwidth]{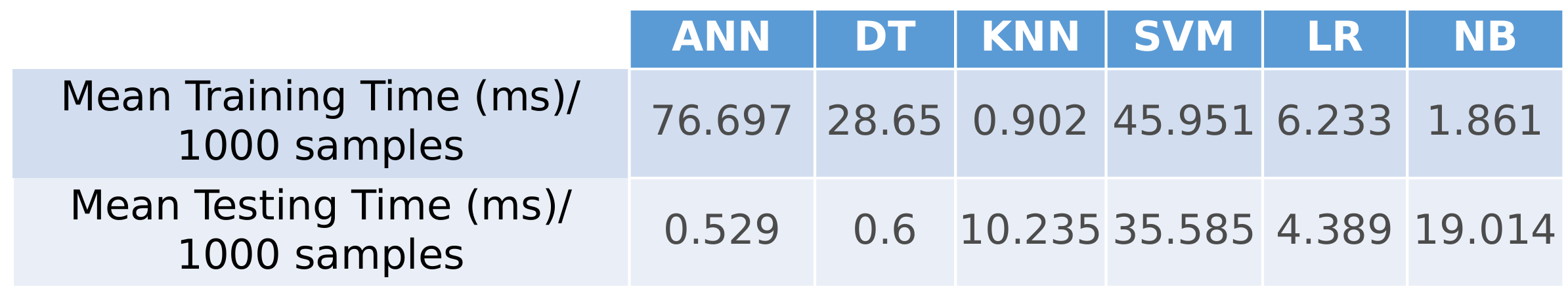}
  \vspace{-0.2 cm}
 \caption{Mean execution time for jamming detection}
\label{fig:detimeresults}
\end{figure}
Additionally, the training time of ANN is longer than other evaluated ML techniques, meanwhile its testing time is the shortest. Using a CPU intel i5-6500, 3.20GHz, we evaluate  the training and testing time for the 6 ML algorithms. Fig. \ref{fig:detimeresults} shows that ANN can take 0.5 ms to evaluate 1000 samples and take 76.6 ms to train the model with 1000 samples. The fast evaluation time of ANN makes it the most suitable for implementation. 

\vspace{-0.2cm}
\subsection{Jamming Attack Localization}
Along with attack detection, it is equally important to locate it. However, locating the sources or access points of attacks is generally complicated and out-of-scope of this paper. We evaluate a single channel attack detection with localization in a network described before.  For the training, we generate 46 samples  for every possible jammed channel by varying the power of the legitimate light sources from 0 to -20 dB and the jamming power which is 1,2 and 3 dB higher.  We generate also a set of authorized samples to make the \textit{Localization Module} aware of the misclassification that can be produced by the \textit{Detection Module}. We use 46 samples for the authorized dataset to have equal number of samples per class which are 33 classes in our case (32 channels + authorized class). The number of data samples in the dataset is 1518. Again, the dataset is divided into training and testing sets with proportion of 75\% and 25\% respectively. 
We evaluate the accuracy of the 6 algorithms with different setups for the attack localization  and analyze TP rate for each channel (i.e., whether an attacked channel is identified correctly or not). More importantly, we use 20 hidden layers for ANN, since lowering the number of hidden layers reduces the localization accuracy. We use a linear kernel for SVM. 
\begin{figure}[t]
 \centering
    \includegraphics[width=0.45\textwidth]{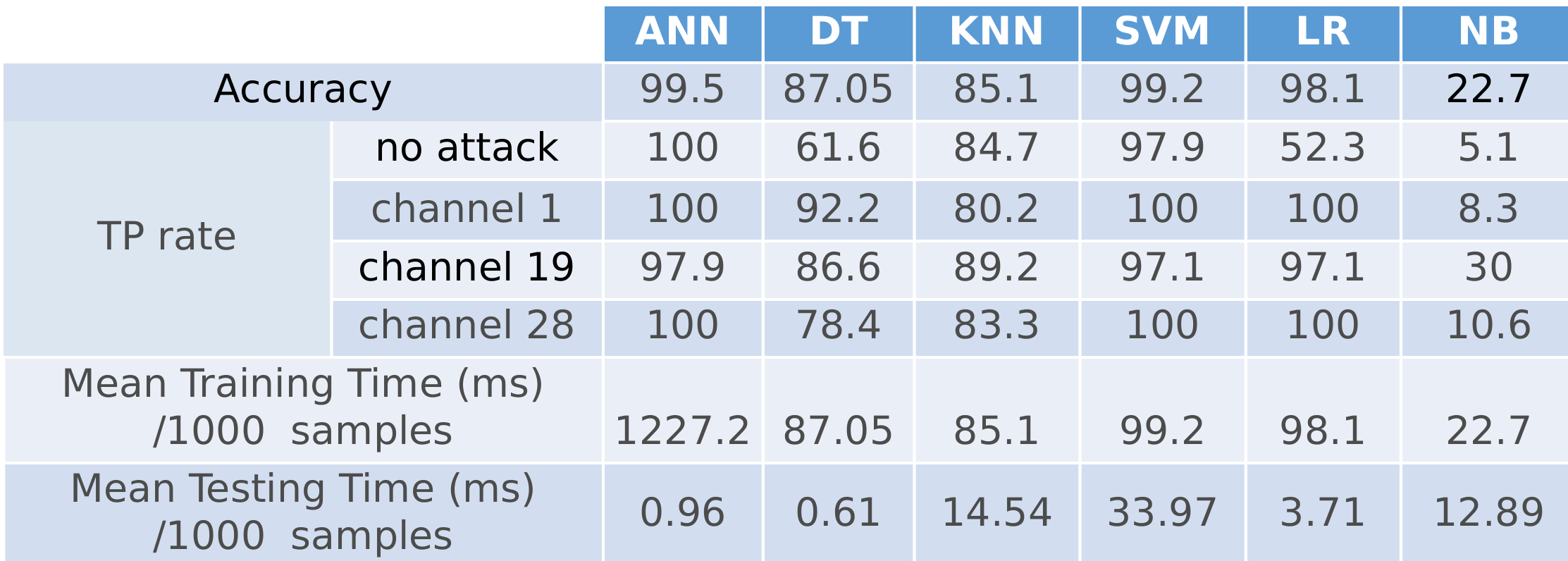}

  \vspace{-0.1 cm}
 \caption{Performance of algorithms for jamming localization}
\label{fig:locresults}
\end{figure}
Confusion matrix is a tool used to analyze the performance of the classifier on every class. We analyzed the confusion matrix of each algorithm and show the main results represented by the TP rate of some classes.  Fig. \ref{fig:locresults} shows the localization accuracy and TP rate of some selected channels using different algorithms. ANN and SVM show the best result of 99\% accuracy and around 99\% TP rate for all channels. LR has an accuracy 98\% while the TP rate for the class "no attack" is 52.3\% which means that when the \textit{Detection Module} misclassify a sample, the \textit{Localization Module} will have only 52.3\% chance to discover this misclassification and 47.7\% chance to alert the Controller that an attack occurred in a wrong channel. DT and KNN have an accuracy around 86\% which is a low accuracy compared to ANN and SVM. NB is a very bad solution which cannot be considered in our problem. 

Based on the accuracy results, it is interesting to compare the time performance of ANN and SVM to be able to decide which method fits better in a real implementation. Fig. \ref{fig:locresults} shows that ANN is much faster than SVM in the inference phase, while it takes more than 10 times longer to be trained. Because the training process is usually running in an off-line mode, it is more important to consider the inference time for the decision which makes ANN again the best option for the localization of a jamming attack being able to evaluate $10^6$ sample per second. 



\vspace{-0.2 cm}
\subsection{A Case Study of Attack Prevention}

Let us assume that the reallocation requests arrive with exponentially distributed mean interarrival time $\tau_r$. If we assume that the holding time of LPs is also exponentially distributed with mean $\tau_c$, then it can be easily shown that the number of arrivals of reallocation requests during the lifetime of an authorized LP is $\frac{\tau_c}{\tau_r}$. However, a reallocation process is only performed with the attack detection probability $p_a$, which means the average number of reconfigurations performed during the lifetime of the LP would reduce to $N_r=\frac{\tau_c}{\tau_r}p_a$. As $N_r$ number of reconfigurations of the LP during its lifetime $\tau_c$ can be divided into $N_r+1$ parts, each part is approximately given as  $\Delta t=\frac{\tau_c}{N_r+1}$ \cite{singh2017combined}. Assuming that one or more attackers can jam or observe authorized LPs  in their proximity, that means to left and right of unauthorized LPs. The amount of successfully jammed data of an authorized LP with a data rate of $b$ bits per unit time in the first interval is $b\Delta t \times p_j$, where $p_j$ is the probability that a LP is jammed in an interval $\Delta t$. Furthermore, in each of the following $N_r$ intervals, the amount of successfully jammed data of the LP is $b\Delta t \times p_j \times p_c$, where $p_c$ is the probability that the LP is jammed  without an interruption at each reallocation. We can obtain these probability terms ($p_j$ and $p_c$) by simulating all (re)allocation instances by an event simulator. Therefore, the amount of data successfully jammed in the LP's lifetime is $b\Delta t\times p_j\left[1+N_r\times p_c\right]$. Moreover, the amount of data carried by the same LP during its lifetime is $b \times \tau_c=b\Delta t (N_r+1)$. Thus, we define a security metric by finding the probability of unsuccessful jamming of an authorized LP, which is given by one minus the ratio of successfully jammed data and total carried data per authorized LP, and it is obtained by Eq. \eqref{eqn:security metric}. 
\begin{equation} \label{eqn:security metric}
\Lambda_j=1- p_j\frac{1+\frac{\tau_c}{\tau_r} p_a  p_c}{1+\frac{\tau_c}{\tau_r}p_a}
\end{equation}

\begin{figure}[t]
 \centering
    \includegraphics[width=0.45\textwidth]{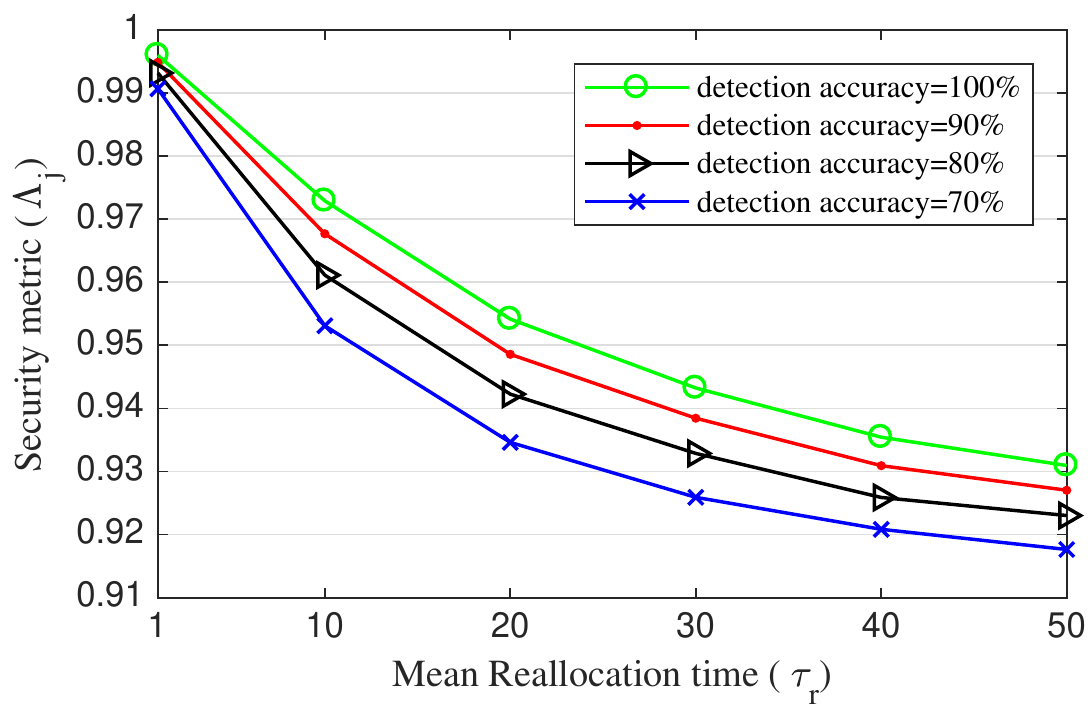}
  \vspace{-0.3 cm}
 \caption{Unsuccessful jamming probability in NSFNET.}
\label{fig:security metric}
\vspace{-0.2 cm}
\end{figure}

A preliminary result of the attack prevention mechanism is shown in Fig. \ref{fig:security metric}, which illustrates the effect of the mean reallocation time ($\tau_r$)  and detection accuracy ($p_a$) on the security of LPs ($\Lambda_j$), as defined by Eq. \eqref{eqn:security metric}. This result is obtained in a well known 14-node NSFNET topology with two shortest paths per node pair, and the number of wavelength channels per link is 80. We generated two types of LP requests, authorized and unauthorized, with a single wavelength demand. The source-destination pair for a  LP request is uniformly selected among all node pairs, and the requests are generated with exponential mean 200 requests per unit time, and their  holding times are exponentially distributed with mean $\tau_c=10$ time units. The reallocation control events are generated with exponential mean $\tau_r$, and it is varied as shown in the horizontal axis in Fig. \ref{fig:security metric}.  We obtained the probability of jamming per reallocation interval ($p_j$) as the ratio of number of jammed interval per authorized LP to the total intervals through a discrete event simulation. Similarly, the probability of jamming without interruption $p_c$ in a reallocation interval is calculated by the ratio of the number of instances where reallocation effect is absent (i.e., a LP is jammed without interruption) to the total reallocation instances.  
In Fig. \ref{fig:security metric}, we see that the data security of LPs against jamming decreases exponentially with the mean reallocation time $\tau_r$.  Furthermore, when the mean reallocation time is  equal to the mean lifetime of authorized LPs, i.e., $\tau_r = \tau_c= 10$, the security decreases only by 3\% for detection accuracy of 100\%, which means even one reconfiguration in the lifetime of a LP is enough to secure 97\% of LPs against  jamming attacks. Therefore, traffic disruption caused by the reallocation scheme is limited by limiting the required number of reallocations per LP. With lower attack detection accuracy and with increasing number of unauthorized LPs in the network, the security against the jamming attacks decreases. 

\begin{figure}[t]
 \centering
    \includegraphics[width=0.45\textwidth]{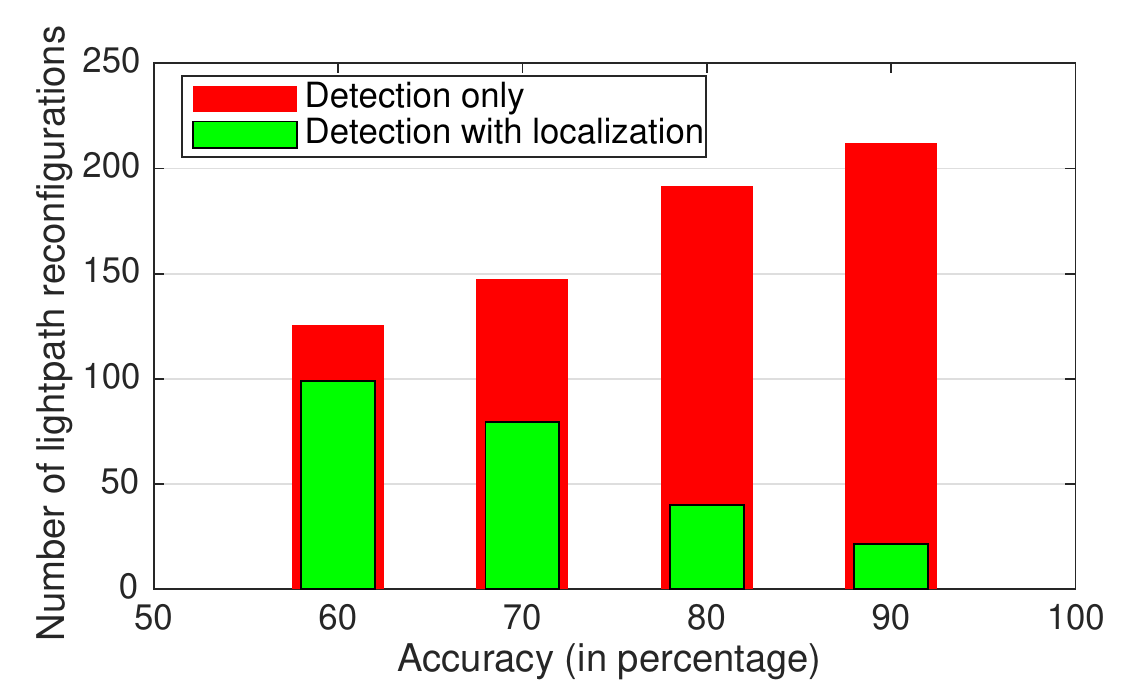}
  \vspace{-0.2 cm}
 \caption{Reconfigurations required per attack detection.}
\label{fig:numberofreconf}
\vspace{-0.2 cm}
\end{figure}

To improve the proposed attack prevention mechanism, we consider the localization accuracy of jamming. The knowledge about the location of an unauthorized LP allows us to just block it as a reactive solution. However, an authorized LP might be blocked due to an error from the \textit{localization module}. To prevent blocking of authorized LPs, we propose to reconfiguring only such authorized  LPs that could be affected by jammed (unauthorized) channels. However, when the \textit{localization module} incorrectly detects  a jammed channel,  reconfiguring the LPs on neighboring channels of an identified unauthorized LPs won't prevent the actual jammed channel, which might exist on another location, from affecting the network. To consider this case, our optimized mechanism reconfigure only the neighboring channels with a probability  $p_l$ and reconfigure all channels with probability $1-p_l$, where $p_l$ is the overall localization accuracy.  
Fig. \ref{fig:numberofreconf} shows the number of reconfigurations required per detection triggered by a controller for attack prevention with and without localization feature. We observe that better the localization accuracy is, less we need to reconfigure LPs. However, reconfigurations increase with the increase in detection accuracy when we don't have the information about the attack's location. The reason is that reconfiguration ($N_r$) increases linearly with the detection accuracy ($P_a$ in Eq. \ref{eqn:security metric}). As a result, with localization, around 8\% of LPs require reconfigurations to secure LPs as compared to the reconfigurations needed using detection only (for 90\% accuracy). 
\section{Conclusion} \label{sec:conclusion}
In this paper, we proposed a ML framework for detecting and preventing jamming attacks in optical networks. We studied the effectiveness of various ML techniques by providing a comparative and detailed analysis using different metrics, such as attack detection accuracy, true positive and negative rates. We find that ANN is the most accurate approach in terms of accuracy and time complexity  to detecting and localizing out-of-band power jamming attacks. Furthermore, we discussed the design of an attack prevention mechanism and studied a security scenario for active jamming attacks by applying a resource reallocation scheme to take preventive actions, and which is shown to improving data security in optical networks. We believe that machine learning-assisted attack detection can be easily integrated in a real optical network, and we plan to further work on a robust attack prevention scheme in future.
\vspace{-0.1 cm}
\bibliographystyle{IEEEtran}
\bibliography{SecurityBib}

\end{document}